\documentstyle[prl,aps]{revtex}

\begin{document}
\author{Jian-Qi Shen$^{1,2}$\footnote{E-mail address: jqshen@coer.zju.edu.cn} and Shao-Long He$^{2}$}
\address{$^{1}$  Centre for Optical
and Electromagnetic Research, Zhejiang University,
Hangzhou Yuquan 310027, P.R. China\\
$^{2}$ Zhejiang Institute of Modern Physics and Department of
Physics, Zhejiang University, \\Hangzhou Yuquan 310027, P.R.
China}
\date{\today}
\title{Aharonov-Carmi effect and energy shift of valency electrons in rotating C$_{60}$ molecules}
\maketitle

\begin{abstract}
It is shown that the valency electrons in rapidly rotating
C$_{60}$ molecules acquire an Aharonov-Carmi (A-C) phase shift,
which is proportional to the molecular angular velocity flux
enclosed by the matter wave of valency electrons on the C$_{60}$
molecular shell. The energy shift of the valency electrons due to
both the molecular rapid rotation and the molecular A-C phase
shift is calculated.

PACS number(s): 03.65.Vf, 61.48.+c
\end{abstract}

Molecular rotation gives rise to some novel effects such as the
quantum decoherence induced by stimulated two-photon Raman
processes, the laser-driving rotational transitions\cite{Adelward}
and the electron spin-rotation geometric phase shift\cite{Shen}.
According to the principle of equivalence, the molecular
rotational dynamics can be treated inside the theoretical
framework of weak gravitational fields. Thus, one can predict some
interesting quantum gravitational (inertial) effects for the
rapidly rotating molecules. Such effects may be particularly
significant for the molecules in solid, since it may provide us
with an insight into the molecular rotational dynamics of
condensed phases\cite{Caccamo}, including the problems of the
intermolecular interaction and phase behavior (phase
diagram)\cite{Caccamo,Zubov,Hasegawa}. As to the quantum
gravitational (inertial) effects, historically, with the
development of laser technology and its application to the
gravitational interferometry experiments\cite{Ahmedov}, some weak
gravitational effects associated with gravitomagnetic fields have
become increasingly important both theoretically and
experimentally, and therefore captured attention of many
investigators\cite{Werner,Atwood,Anandan}. During the past 20
years, neutron interferometry was developed with an increasing
accuracy. For example, by using these technologies, Werner {\it et
al.} investigated the neutron analog of the
Foucault-Michelson-Gale effect in 1979\cite{Werner} and Atwood
{\it et al.} found the neutron Sagnac effect in 1984\cite{Atwood}.
Aharonov, Carmi {\it et al.}\cite{Anandan,Carmi} proposed a
gravitational analog to the Aharonov-Bohm effect\cite{Bohm}. This
is a geometric (topological) effect of vector potential of
gravity: specifically, in a rotating frame the matter wave
propagating along a closed path will acquire a nonintegral phase
(geometric phase). This phenomenon has now been called the
Aharonov-Carmi (A-C) effect, or the gravitational (inertial)
Aharonov-Bohm effect. Overhauser, Colella\cite{Overhauser}, Werner
and Standenmann {\it et al.}\cite {Werner} have demonstrated the
existence of such a geometric effect by means of the
neutron-gravity interferometry experiments. All these phenomena
were found and studied in macroscopic experiments. In the present
report, we will consider the molecular A-C effect in rotating
C$_{60}$ molecules. Here the C$_{60}$ rotation provides an
effective weak gravitomagnetic field.

In the literature, the nuclear-magnetic resonance (NMR)
studies\cite{Yannoni} and the quasielastic neutron scattering
experiments\cite{Neumann} were performed to obtain the information
on the rotational dynamics of C$_{60}$ molecules in solid phases.
It was shown by these methods that C$_{60}$ molecules in the
orientationally disordered phase experience a rapid rotation, the
rotational correlation time of which may be picoseconds. It
follows that the molecular dynamics of C$_{60}$ rotation is of
particular importance, since it has close relation to the
molecular thermal motion, phase transition and crystal structure
of C$_{60}$ solid\cite{Caccamo,Zubov,Hasegawa}.

As for the electronic structure and bonding in C$_{60}$ molecule,
the H\"{u}ckel molecular orbital theory for non-planar conjugated
organic molecules has been applied to study the electronic state
and properties of the icosahedral geometry of
C$_{60}$\cite{Haddon}. The molecule was calculated to have a
stable closed shell singlet ground electronic state. Although
there are two independent bond types ({\it i.e.}, 30 bonds which
lie solely in 6-MRs, while 60 bonds form the edges of both a 5-
and a 6-MR), the experimental observations showed that the
molecule is non-alternant, but that the charge densities are all
equal as a result of the symmetry of the molecule
itself\cite{Haddon,Disch}. In order to study the molecular A-C
effect in C$_{60}$, here we consider only the interference
behavior of one of the valency electrons ({\it e.g.}, the
delocalized $\pi$-bond electrons on shell), which undergo an
inertial coupling ({\it i.e.}, the Coriolis force) to the
molecular rotation. The Hamiltonian of the valency electron that
is acted upon by a Coriolis force due to the molecular rapid
rotation is of the form $H({\bf r})=\left(-i{\hbar}\nabla-2m_{\rm
e}{\bf a}\right)^{2}/(2m_{\rm e})$, where $m_{\rm e}$ denotes the
electron mass, and the effective gravitomagnetic vector potential
${\bf a}$ is so defined that the molecular angular velocity
$\vec{\omega}=\nabla\times {\bf a}$. In general, ${\bf a}$ in a
spherical coordinate system ($r, \theta, \varphi$) can be written
as $a_{r}=0$, $a_{\theta}=0$ and $a_{\varphi}=(\omega r/2)\sin
\theta$. The eigenvalue equation of $H$ is $H({\bf r})\Psi({\bf
r})=E\Psi({\bf r})$. It is apparently seen that the
stationary-state wavefunction $\Psi({\bf r})$ of the electron can
be rewritten as
\begin{equation}
\Psi({\bf r})=\exp\left(\frac{i2m_{\rm e}}{\hbar}\int^{\bf
r}_{{\bf r}_{0}} {\bf a}\cdot {\rm d}{\bf l}\right)\Psi_{0}({\bf
r}),      \label{eq2}
\end{equation}
where ${\bf r}_{0}$ denotes the initial position vector of the
integral range. The wavefunction $\Psi_{0}({\bf r})$ satisfies the
stationary equation $ -\left({\hbar^{2}}/{2m_{\rm
e}}\right)\nabla^{2}\Psi_{0}({\bf r})=E\Psi_{0}({\bf r}) $. Thus,
the A-C phase shift acquired by the valency electron is $\Delta
\phi=({2m_{\rm e}}/{\hbar})\oint_{\bf l} {\bf a}\cdot {\rm d}{\bf
l}$, which can be rewritten as
\begin{equation}
\Delta \phi=\frac{2m_{\rm
e}}{\hbar}\mathop{{\int\!\!\!\!\!\int}\mkern-21mu
\bigcirc}\limits_{\bf S}\left(\nabla\times {\bf a}\right)\cdot{\rm
d}{\bf S}=\frac{2m_{\rm e}}{\hbar}\vec{\omega}\cdot{\bf A}
\label{ACphase}
\end{equation}
with ${\bf A}$ being the area vector circulated by the standing
wave of the valency electron on the C$_{60}$ molecular shell.
Obviously, the expression $\vec{\omega}\cdot{\bf A}$ in Eq.
(\ref{ACphase}) is just the molecular rotational angular velocity
flux. According to the principle of equivalence, the Coriolis
force acting upon the valency electron can be considered a
gravitomagnetic Lorentz force. In this sense, the expression
$\vec{\omega}\cdot{\bf A}$ can also be referred to as the
gravitomagnetic flux enclosed by the matter wave of the valency
electron. In what follows, we will evaluate the order of magnitude
of the A-C phase shift acquired by the electron. In the
high-temperature phase (orientationally disordered phase),
$\omega$ may be $10^{11}$ rad/s\cite{Johnson}. Thus the order of
magnitude of the molecular A-C phase shift in a rotating C$_{60}$
molecule may be $10^{-3}$. Here we have taken a typical value
$3\times 10^{-19}$ m$^{2}$ for the modulus of the area vector
${\bf A}$. Since it has no connection with the dynamical
quantities such as electron energy and velocity, the A-C phase
shift $\Delta \phi$ is viewed as a geometric (topological) phase
shift. It should be noted that such an additional phase shift will
have influence on the wave interference (to form the standing
wave) of valency electrons on the C$_{60}$ molecular spherical
shell. As a result, the energy of the valency electrons in
C$_{60}$ molecules will inevitably be shifted, which may have an
observable effect on the C$_{60}$ photoelectron spectroscopy. In
the following, we will discuss such an energy shift resulting from
the C$_{60}$ molecular A-C effect.

In the spherical coordinate system, the on-shell wavefunction,
$\Psi_{0}({\bf r})$, of the valency electron can be written in the
form $R(r_{0})\Theta(\theta)\Phi(\varphi)$, where $r_{0}$ denotes
the radius of the C$_{60}$ molecule. With the help of the
eigenvalue equation of $\Psi_{0}({\bf r})$, one can arrive at
\begin{equation}
\left\{
\begin{array}{ll}
& E=\lambda\frac{\hbar^{2}}{2m_{\rm e}r_{0}^{2}},    \\
&  \frac{1}{\sin \theta}\frac{\rm d}{{\rm d}\theta}\left(\sin \theta\frac{{\rm d}\Theta}{{\rm d}\theta}\right)
+\left(\lambda-\frac{m^{2}}{\sin^{2}\theta}\right)\Theta =0,    \\
& \Phi(\varphi)=\frac{1}{\sqrt{2\pi}}\exp (im\varphi).
\end{array}
\right. \label{set}
\end{equation}
Thus, it follows from Eqs. (\ref{eq2}) and (\ref{set}) that the
$\varphi$-dependent total phase in $\Psi({\bf r})$ is
\begin{equation}
\phi(\varphi)=m\varphi+\frac{2m_{\rm e}}{\hbar}\int^{\bf r}_{{\bf
r}_{0}} {\bf a}\cdot {\rm d}{\bf l}=m'\varphi      \label{total}
\end{equation}
with
\begin{equation}
m'=m+\frac{m_{\rm e}\omega r_{0}^{2}\sin^{2}\theta}{\hbar}.
\label{eqeq6}
\end{equation}
Note that the expressions $a_{r}=0$, $a_{\theta}=0$ and
$a_{\varphi}=(\omega r/2)\sin \theta$ for the gravitomagnetic
vector potential ${\bf a}$ have been inserted into the total phase
(\ref{total}). Since the wavefunction has to be a single-valued
function of position, $m'$ in Eqs. (\ref{total}) and (\ref{eqeq6})
must be an integer. This, therefore, implies that $m$ in Eqs.
(\ref{set}) is no loner an integer. Instead, as seen from
 Eq. (\ref{eqeq6}), it is a function of $\theta$. Inserting the
relation $m=m'-{m_{\rm e}\omega r_{0}^{2}\sin^{2}\theta}/{\hbar}$
into the second equation of Eqs. (\ref{set}), one can obtain
\begin{equation}
\frac{1}{\sin \theta}\frac{\rm d}{{\rm d}\theta}\left(\sin
\theta\frac{{\rm d}\Theta}{{\rm
d}\theta}\right)+\left(\lambda'-\frac{m'^{2}}{\sin^{2}\theta}\right)\Theta
=0,     \label{eq7}
\end{equation}
where
\begin{equation}
\lambda'=\lambda+\frac{2m_{\rm e}m'\omega r_{0}^{2}}{\hbar}.
\label{eq8}
\end{equation}
Note that in Eq. (\ref{eq7}), we have ignored the small term
$\left(m_{\rm e}\omega
r_{0}^{2}/\hbar\right)^{2}\sin^{2}{\theta}\Theta$, which is about
only one part in $10^{6}$ of the retained term
$(m'^{2}/\sin^{2}\theta)\Theta$. It is clear that the parameter
$\lambda'$ in Eq. (\ref{eq7}) takes $\lambda'=l(l+1)$, where $l$
is the angular quantum number (integer). Thus, it follows from the
expression (\ref{eq8}) that $\lambda$ is constant (independent of
position $\theta$) but no loner an integer. Substitution of the
expression $\lambda=\lambda'-{2m_{\rm e}m'\omega
r_{0}^{2}}/{\hbar}$ into the energy eigenvalue $E$ in Eqs.
(\ref{set}), one can arrive at the expression for the energy shift
of valency electron, {\it i.e.},
\begin{equation}
\Delta E=-m'\hbar\omega.     \label{eq9}
\end{equation}
This is the energy-shift effect of the valency electron due to the
C$_{60}$ molecular rotation (and hence the molecular A-C effect).
The physical meanings of (\ref{eq9}) can be considered an
interaction between the gravitomagnetic moment ($m'\hbar$) and the
gravitomagnetic field ($\omega$). In a orientationally disordered
phase, such an energy shift may be of the order of magnitude of
$10^{-4}\sim 10^{-3}$ eV.

Since C$_{60}$ molecules possess a spherically symmetric geometric
shape, it rotates rapidly at about $10^{9}$ rad/s (in the
orientationally ordered phase) and at about $10^{11}$ rad/s (in
the orientationally disordered phase)\cite{Johnson}. Moreover, in
C$_{60}$ condensed phases the molecule is subject to a fluctuation
(precession and nutation) of the rotational motion caused by the
noncentral intermolecular potential. It can be readily verified
that the precessional frequency of C$_{60}$ rotation in the
orientationally ordered phase may be very large, {\it e.g.}, it
ranges from $10^{12}$ to $10^{14}$ rad/s\cite{Shen}. Here it is
worth noting that even though the molecular precessional frequency
is so great, the result of (\ref{eq9}) still holds, since the
noncentral intermolecular potential leads only to the variation of
the direction of the angular velocity of rotating C$_{60}$
molecules and the magnitude of angular velocity does not alter
much. It is known that the analysis of the intermolecular
potential is essential for both the molecular-dynamics (MD)
simulation and the study of the high-temperature phase diagram of
C$_{60}$ solid\cite{Caccamo,Zubov,Hasegawa}. So, the treatment of
the fluctuation of the molecular rotational motion is also
important for considering these subjects. Because the electron
spin can be coupled to the molecular angular velocity
(spin-rotation coupling)\cite{Mashhoon}, such a significant
precession of C$_{60}$ rotation will unavoidably result in a
so-called spin-rotation phase shift acquired by the valency
electrons\cite{Shen}. Thus, both the molecular rotation and the
precession can yield the valency-electron state plus the geometric
phase shifts. We think that the information on these geometric
phase shifts (and hence energy shift) may be read off from the
photoelectron spectroscopy of C$_{60}$. For these reasons, the
molecular A-C effect, the consequent valency-electron energy shift
and some properties relevant to the above-mentioned effects
deserve consideration for the investigation of the molecular
rotational dynamics of C$_{60}$ solid.

\textbf{Acknowledgements} This work was supported in part by the
National Natural Science Foundation of China under Project No.
$10074053$ and the Zhejiang Provincial Natural Science Foundation
under Project No. $100019$.

\end{document}